\PassOptionsToPackage{prologue,dvipsnames,table}{xcolor}

\documentclass[conference]{IEEEtran}
\IEEEoverridecommandlockouts

\usepackage{etc}
\usepackage{algorithm}
\usepackage{algorithmic}
\usepackage{url}
\usepackage{multirow}
\usepackage{balance}
\usepackage{float}

\usepackage[table]{xcolor}
\usepackage{array}
\usepackage{booktabs}
\usepackage{graphicx} 
\usepackage{siunitx} 

\definecolor{ctrain}{HTML}{C3C31E}

\definecolor{cinfer}{HTML}{b573f8}      
\definecolor{csystem}{HTML}{f873b5}     
\definecolor{cdata}{HTML}{f873f8}       
\definecolor{carch}{HTML}{626281}       

\newcommand{\datacell}{\raisebox{-0.5ex}{\textcolor{cdata}{\rule{1.5ex}{1.5ex}}}}
\newcommand{\modelcell}{\raisebox{-0.5ex}{\textcolor{carch}{\rule{1.5ex}{1.5ex}}}}
\newcommand{\traincell}{\raisebox{-0.5ex}{\textcolor{ctrain}{\rule{1.5ex}{1.5ex}}}}
\newcommand{\systemcell}{\raisebox{-0.5ex}{\textcolor{csystem}{\rule{1.5ex}{1.5ex}}}}
\newcommand{\infercell}{\raisebox{-0.5ex}{\textcolor{cinfer}{\rule{1.5ex}{1.5ex}}}}
\newcommand{\emptycell}{\raisebox{-0.5ex}{\textcolor{gray!20}{\rule{1.5ex}{1.5ex}}}}
\newcommand{\vindex}[1]{$V_{#1}$}

\newcommand{\stages}[5]{%
\makebox[7.5ex][c]{%
#1\hspace{0.2ex}%
#2\hspace{0.2ex}%
#3\hspace{0.2ex}%
#4\hspace{0.2ex}%
#5%
}}

\usepackage{tabularx}
\newcolumntype{C}{>{\centering\arraybackslash}p{0.045\textwidth}}   
\newcolumntype{Z}{>{\raggedright\arraybackslash}X}                  

\newcolumntype{N}{>{\centering\arraybackslash}X}

\newcolumntype{D}{>{\raggedright\arraybackslash}p{0.2\textwidth}}

\newtcolorbox{rqbox}[1][]{
    colback=green!10,
    colframe=gray,
    arc=1mm,
    boxrule=0.5pt,
    coltitle=black,
    fonttitle=\bfseries,
    title=#1,
    top=0.5mm,
    bottom=0.5mm,
    left=1mm,
    right=1mm,
    toptitle=0.5mm,
    bottomtitle=0.5mm
}

\def\BibTeX{{\rm B\kern-.05em{\sc i\kern-.025em b}\kern-.08em
    T\kern-.1667em\lower.7ex\hbox{E}\kern-.125emX}}

\begin{document}

\title{Tu(r)ning AI Green: Exploring Energy Efficiency Cascading with Orthogonal Optimizations}

\author{
\IEEEauthorblockN{Saurabhsingh Rajput, Mootez Saad, Tushar Sharma}
\IEEEauthorblockA{\textit{Dalhousie University} \\
Halifax, Canada \\
\{saurabh,mootez,tushar\}@dal.ca}}

\maketitle
\thispagestyle{plain}
\pagestyle{plain}

\begin{abstract}
\aim{}'s exponential growth intensifies computational demands and energy challenges. While practitioners employ various optimization techniques,
that we refer as ``knobs'' in this paper, 
to tune model efficiency, these are typically afterthoughts and reactive ad-hoc changes applied in isolation without understanding their combinatorial effects on energy efficiency.
This paper emphasizes on treating energy efficiency as the first-class citizen and as a fundamental design consideration for a compute-intensive pipeline. We show that strategic selection across five AI pipeline phases (data, model, training, system, inference) creates cascading efficiency. Experimental validation shows orthogonal combinations reduce energy consumption by up to $94.6$\% while preserving $95.95$\% of the original F1 score of non-optimized pipelines. This curated approach provides actionable frameworks for informed sustainable \aim{} that balance efficiency, performance, and environmental responsibility.
\end{abstract}


\section{Introduction}
The exponential growth of generative Artificial Intelligence (\aim{}) has fundamentally transformed modern technology landscapes, with its market size growing at an extraordinary rate of $27.67$\% annually~\cite{statistaArtificialIntelligence}. 
Software Engineering (\textsc{se}) stands at the forefront of this \aim{} revolution, integrating \aim{} into critical tasks from code completion to code generation and to testing and deployment.
While this advancement drives innovation, it comes with a hidden cost. Modern \aim{} models consume enormous computational resources, resulting in significant energy usage and carbon emissions.
In response, \textcolor{ForestGreen}{\textbf{Green \aim{}}} initiatives advocate making energy and compute efficiency as the core evaluation criteria for \aim{} research, alongside traditional metrics such as accuracy.
These initiatives focus on developing energy-efficient algorithms, optimizing model architectures, and reducing computational requirements.

Practitioners now employ numerous optimization techniques such as distillation,
pruning,
and quantization to
enhance efficiency. However, these approaches are typically applied in isolation, without a systematic understanding of how they influence the overall performance metrics and energy efficiency when combined. Current optimization efforts mainly focus only on inference and training stages, often implemented as reactive measures after significant energy expenditure. This fragmented application overlooks potential synergies across the development lifecycle \textit{and treats energy impact as a secondary concern}. 

This challenge is compounded by the continued increase in aggregated energy consumption despite efficiency 
gains. Newer language models demonstrate remarkable parameter efficiency improvements, yet their collective energy footprint expands exponentially due to increased adoption. This paradox underscores the need for an integrated, software-engineering–driven agenda. Our work answers this call, aligning with Cruz \etal{}'s roadmap~\cite{cruz2025greeningaienabledsystemssoftware} to guide the development of truly sustainable \aim{} systems.

In this work, we demonstrate how strategic usage 
of Green \aim{} optimization techniques across five development phases---data, model architecture, training, system design, and inference---creates cascading efficiency gains. 
Through rigorous experimentation, we show that orthogonal  combinations (those targeting distinct resource constraints) of such optimization techniques  yield compounded benefits, reducing energy consumption between $4.6$\% and $94.6$\% while maintaining model performance ($-4.05$ to $+0.12$\%) in terms of F1 score. The study emphasizes transforming energy efficiency from an afterthought to a core design consideration, empowering practitioners to achieve sustainable \aim{} without compromising the capability of their \aim{} pipeline.

\section{The Optimization-Efficiency Conundrum}\label{status_quo}

The computational intensity of state-of-the-art \aim{} models continues to escalate with training requirements doubling approximately every $3.4$ months\textemdash outpacing hardware efficiency gains~\cite{Perrault2019}. This acceleration is evident in model scaling from \gpt{}-$3$'s $175$ billion parameters to PaLM's $540$ billion and beyond, 
with each leap increasing energy demands. 

\subsection{Where Current Practice Falls Short}

\begin{figure*}[ht]
    \centering
    \includegraphics[width=\textwidth]{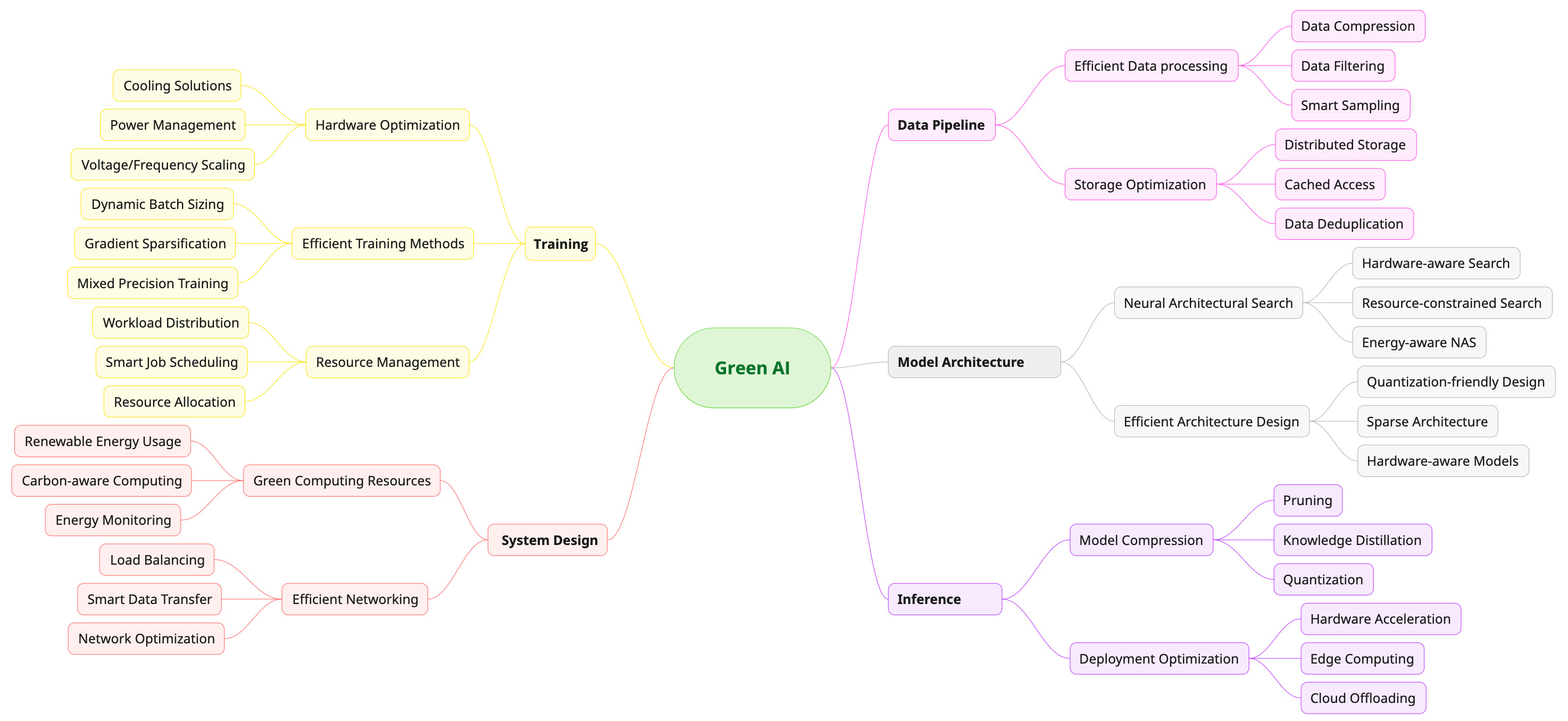}
    \caption{Green \aim{} techniques: A holistic view of sustainable \aim{} techniques across different development phases.  }
    \label{fig:green_ai_mindmap}
\end{figure*}

The high rate of energy consumption despite the availability of techniques that make \aim{} models computationally efficient is attributed to the \textit{compute-centric} nature of such methods, where practitioners prioritize hardware utilization reduction and 
accuracy metrics while neglecting energy implications. 
As shown in Figure~\ref{fig:green_ai_mindmap}, 
numerous optimization \textit{knobs} are proposed\textemdash knobs are a collection of configuration parameters controlled by a corresponding technique or approach  that impacts the energy consumption in an \aim{} pipeline. However, developers deploy them reactively as an afterthought.
This approach highlights two critical aspects. First, energy consumption constitutes a fundamental optimization dimension that needs to be treated as a first-class citizen similar to accuracy and other performance metrics.
Second, though these knobs may individually lead to some energy efficiency, their careful combinations may exhibit orthogonal effects that may lead to significant efficiency gain.

\subsection{Green Practices as First-class Citizens
}\label{green_ai_pipeline}

Most optimization techniques deployed today originate from \textit{compute-centric} 
experiments where energy efficiency is not considered as an objective.
\textcolor{ForestGreen}{\textbf{Our vision emphasizes energy consumption as a fundamental design constraint across all development phases, enabling practitioners to \textit{tune their AI green} through strategic combinations of orthogonal optimization knobs.}} We propose transforming sustainability from an afterthought to a core design principle by:
\begin{enumerate}
    \item embedding energy awareness during \textit{design, development, and execution} of AI pipelines,
    \item combining knobs \textit{orthogonally} for amplified efficiency, and 
    \item explicitly balancing accuracy-energy tradeoffs.
\end{enumerate}

The following sections detail how implementing this vision across five critical development phases 
(shown in Figure~\ref{fig:green_ai_mindmap})
creates cumulative
energy savings while maintaining performance.

\subsubsection*{\textcolor{cdata}{\textbf{Data pipeline}}} 

Current data practices emphasize quantity over quality, with companies amassing enormous datasets by scraping the Internet. Such web-scale data collection often leads to redundant storage and repeated processing of similar content. Modern \llms{} are trained on trillions of tokens; for instance, \textsc{llama-3}
was trained on approximately 15 trillion tokens. Despite diminishing returns in model performance as datasets grow,
the industry trend continues to favor larger datasets~\cite{cheng2024dateddatatracingknowledge}.

Effective data management practices must shift from merely optimizing storage efficiency to a more resource-aware approach. Data preprocessing lays the groundwork for model performance, influencing resource utilization. Instruction-tuning \llms{} on curated, high-quality prompts can yield performance comparable or superior to models trained on larger datasets~\cite{zhou2024lima}. Techniques such as smart-sampling, model input trimming, efficient tokenization, and strategic-filtering
can reduce computational loads without sacrificing accuracy. Energy-conscious data pipelines
can effectively balance computational and storage trade-offs. Additionally, smart sampling strategies and data deduplication methods minimize redundant processing, leveraging heuristics to eliminate low-quality data early and thus reduce computational overhead. Locality-aware distributed storage solutions further mitigate unnecessary data transfers, while automated quality assessment frameworks prevent processing redundant or irrelevant data. Finally, efficient data versioning and caching mechanisms help minimize redundant computation and storage overhead, reinforcing an energy-aware data handling pipeline.

\subsubsection*{\textcolor{carch}{\textbf{Model architecture}}}  

Current architectural trends favor larger models with billions of parameters, following the \textit{bigger is better} scaling law~\cite{ibmBiggerLanguage}. The intent is to prioritize performance gains through increased model capacity and computation. 

This phase can be benefited by using modern techniques focusing on efficient model architectures. This includes sparse architectures \eg{} pruned models, reducing parameter counts while maintaining model capacity, using activation functions such as ReLU where applicable that minimize computation and memory access, and adopting hardware-aware designs such as flash attention
that optimize performance for specific hardware. Additionally, quantization-friendly architectures and energy-aware Neural Architecture Search (\textsc{nas})
can systematically guide the development of more sustainable architectures.

\subsubsection*{\textcolor{ctrain}{\textbf{Training phase}}}

Contemporary training practices focus on achieving state-of-the-art performance through computionally-intensive approaches. Models are typically trained on massive \textsc{gpu} clusters, with repeated training runs to tune hyperparameters. 
For example, \gpt{}-$4$'s training estimated to consume over $2 \times 10^{10}$ peta\textsc{flops}~\cite{gpt4compute}. 
This phase emphasizes achieving maximum accuracy, often through brute-force approaches that consume substantial energy.

At the computational level, mixed precision training significantly reduces both memory requirements by utilizing lower numerical precision where possible~\cite{micikevicius2017mixed}.
You~\etal{}~\cite{you2023zeus} demonstrated that dynamic batch sizing can reduce energy consumption by up to $75.8$\% through adaptive resource allocation.
For resource-constrained environments, gradient sparsification \ie{} omitting gradient components below a certain threshold, offers efficiency by updating only the most significant model parameters~\cite{10026255}. 

\aim{} training must evolve from brute-force computation to intelligent and optimal resource utilization. This transformation would require novel adaptive resource allocation techniques that will dynamically scale based on model complexity and task criticality. Distributed training frameworks must integrate renewable energy awareness and carbon footprint optimization at their core. Smart batch sizing strategies should maintain convergence while respecting energy bounds. Energy-aware hyperparameter optimization should consider computational costs alongside performance metrics. Newer generations of optimizers and loss functions must incorporate energy consumption as an explicit optimization target, balancing model performance with computational efficiency. Organizational collaboration in model training should become essential to eliminate redundant computation and resource waste. Additionally, code language models should be pretrained with energy efficiency patterns as a target objective, or fine-tuned using reinforcement learning to generate code that optimizes both functional correctness and energy efficiency.

\subsubsection*{\textcolor{csystem}{\textbf{System design}}} 
Current system architectures for \aim{}-enabled applications prioritize performance and compute scaling, treating energy efficiency as an afterthought. Techniques such as load balancing focus on throughput and reliability but largely ignore energy consumption. AI training jobs are queued based on requested compute resources, and autoscalers provision capacity using only CPU and RAM utilization metrics—without power monitoring, automated resource allocation, or carbon-aware scheduling. Moreover, heterogeneous hardware efficiency profiles are treated uniformly, and the absence of standardized energy-monitoring systems or open data on consumption creates significant barriers to sustainable computing.

Making the development of \aim{} models more energy-aware  
requires integrating sustainability deeply into system architectures. This includes adopting carbon-aware scheduling to align workloads with renewable energy availability, comprehensive monitoring tools for real-time resource tracking, and leveraging predictive analytics for enhanced resource management, such as power capping and frequency scaling. Collaborative frameworks enabling shared infrastructure are essential to facilitate optimal resource utilization, aligning system performance effectively with sustainability objectives.

\subsubsection*{\textcolor{cinfer}{\textbf{Inference phase}}}

Current inference deployments predominantly prioritize low latency and high throughput, often neglecting energy efficiency. Large-scale deployments frequently replicate model instances across multiple global data centers to ensure rapid response times, further amplifying resource consumption. 
This trend is particularly pronounced with larger models like \gpt{}-4, whose inference queries demand substantial computational resources, exacerbating operational energy footprints.
Since inference workloads constitute approximately 80\% of total \aim{} computational demands in data centers, compared to training's 20\%~\cite{seDisruptionChallenges}, post-training optimization becomes crucial for sustainable operations.
Techniques such as dynamic pruning,
quantization,
knowledge distillation,
and KV caching
significantly reduce inference-related energy usage.

Moving forward, inference optimization must balance between resource-conscious and latency-focused.
Employing context-sensitive model selection, including mixtures-of-experts, can dynamically scale computational resources according to specific task requirements. Intelligent pruning approaches tailored to deployment constraints, shared inference infrastructure, and real-time energy monitoring systems equipped with automated optimization capabilities will effectively align computational performance with broader sustainability objectives.

\begin{rqbox}
\textbf{Takeaways.} AI development treats energy as an after-thought: each stage—data collection, model design, training, deployment, and inference—optimizes for accuracy, speed, or scale first, then applies scattered tweaks to save compute because Because energy is not a front-line design constraint, these scattered fixes fail to control the overall power surge.  
Embedding energy as a first-class objective from the outset ensures that early decisions propagate efficiency instead of inflating demand.
\end{rqbox}

\section{Cascading Effects Across the Pipeline}
Our experiments aim to bridge the gap between theoretical Green \aim{} principles and practical implementation, to assess their impact. 
Our aim is to reveal the difference between the fragmented and ad-hoc optimization approach and systematically considering energy efficiency as a first-class citizen. 

We evaluate each optimization technique, \ie{} \textit{knob}, across two critical dimensions: (1) \textit{knob isolation}, to quantify the efficacy of individual Green AI techniques, and (2) \textit{knob combination}, to uncover synergistic, potentially cascading effects. 
This analysis validates our central hypothesis that if optimizations are strategically curated, they deliver cascaded 
benefits; otherwise, poorly curated optimizations may yield diminished or even canceling effects.

\begin{table*}[ht]
\centering
\caption{Comprehensive Analysis of Optimization Variants. \infercell: Inference, \datacell: Data, \systemcell: System, \modelcell: Model, \traincell: Train}
\label{tab:comprehensive-analysis}
\footnotesize
\setlength{\tabcolsep}{2pt}
\begin{tabularx}{\textwidth}{@{}
  l            
  D            
  c            
  *{7}{N}      
  N            
  N            
  *{2}{N}      
  N            
@{}}
\toprule
& & & \multicolumn{8}{c}{\textbf{Energy} \scriptsize(\si{\kWh})} & 
\textbf{Perf.} & \multicolumn{3}{c}{\textbf{Time} \scriptsize(\si{\second})}\\
\cmidrule(lr){4-11}\cmidrule(lr){12-12}\cmidrule(lr){13-15}
\raisebox{1.5ex}{\textbf{Variant}} &
\raisebox{1.5ex}{\textbf{Description}} &
\raisebox{1.5ex}{\textbf{Stage(s)}} &

\rotatebox{0}{\parbox{0.045\textwidth}{\centering Data\\\scriptsize($\times10^{-5}$)}} &
\rotatebox{0}{\parbox{0.045\textwidth}{\centering Token\\\scriptsize($\times10^{-6}$)}} &
\rotatebox{0}{\parbox{0.045\textwidth}{\centering Load\\\scriptsize($\times10^{-5}$)}} &
\rotatebox{0}{\parbox{0.045\textwidth}{\centering Train\\\scriptsize($\times10^{-1}$)}} &
\rotatebox{0}{\parbox{0.045\textwidth}{\centering Save\\\scriptsize($\times10^{-5}$)}} &
\rotatebox{0}{\parbox{0.045\textwidth}{\centering Eval\\\scriptsize($\times10^{-2}$)}} &
\rotatebox{0}{\parbox{0.045\textwidth}{\centering Total\\\scriptsize($\times10^{-1}$)}} &
\rotatebox{0}{\parbox{0.045\textwidth}{\centering $\Delta$Total\\\scriptsize($\%$)}} &
\rotatebox{0}{\parbox{0.045\textwidth}{\centering $\Delta$F1\\\scriptsize($\%$)}} &
\rotatebox{0}{\parbox{0.045\textwidth}{\centering Total\\\scriptsize($\times10^{3}$)}} &
\rotatebox{0}{\parbox{0.045\textwidth}{\centering Eval\\\scriptsize($\times10^{2}$)}} &
\rotatebox{0}{\parbox{0.045\textwidth}{\centering $\Delta$Eval\\\scriptsize($\%$)}} \\
\midrule
$V_{0}$  & Baseline (ModernBERT) & \stages{\emptycell}{\emptycell}{\emptycell}{\emptycell}{\emptycell}
     & $2.79$ & $4.91$ & $2.47$ & $4.68$ & $3.84$ & $4.44$ & $5.12$
     & $0.0$ & $+0.00$ & $7.92$ & $5.17$ & $+0.0$ \\
\midrule
$V_{1}$  & Gradient Checkpointing & \stages{\emptycell}{\emptycell}{\traincell}{\emptycell}{\emptycell}
     & \colorbox{SpringGreen}{$2.68$} & $4.51$ & $2.55$ & \colorbox{Salmon}{$6.25$} & $4.03$ & $4.45$ & \colorbox{Salmon}{$6.70$}
     & $+30.7$ & $-0.03$ & $10.4$ & $5.18$ & $+0.1$ \\

$V_{2}$  & LoRA PEFT & \stages{\emptycell}{\modelcell}{\traincell}{\emptycell}{\infercell}
     & $3.29$ & \colorbox{SpringGreen}{$3.76$} & $3.52$ & \colorbox{SpringGreen}{$0.59$} & $1.21$ & $1.62$ & $0.80$
     & $-84.3$ & $-4.00$ & $1.7$ & $2.35$ & $-54.5$ \\

$V_{3}$  & Quantization & \stages{\emptycell}{\modelcell}{\traincell}{\systemcell}{\infercell}
     & $4.75$ & $542$ & \colorbox{Salmon}{$5.22$} & $4.41$ & \colorbox{SpringGreen}{$0.76$} & $4.38$ & $4.85$
     & $-5.3$ & $-0.06$ & $11.6$ & $6.43$ & $+24.4$ \\

$V_{4}$  & Tokenizer Optimization & \stages{\datacell}{\emptycell}{\emptycell}{\systemcell}{\infercell}
     & $3.19$ & $4.79$ & $2.40$ & $4.68$ & $3.68$ & $4.45$ & $5.13$
     & $+0.1$ & $+0.11$ & $7.93$ & $5.19$ & $+0.3$ \\

$V_{5}$  & Power-Limit (100 W) & \stages{\emptycell}{\emptycell}{\emptycell}{\systemcell}{\emptycell}
     & $7.86$ & $5.01$ & $2.42$ & $5.92$ & $3.47$ & \colorbox{Salmon}{$5.56$} & $6.48$
     & $+26.5$ & $-0.05$ & $20.5$ & $13.8$ & $+167.6$ \\

$V_{6}$  & Optimizer Tuning & \stages{\emptycell}{\emptycell}{\traincell}{\emptycell}{\emptycell}
     & $4.88$ & $4.57$ & $2.40$ & $4.44$ & $3.72$ & $4.45$ & $4.89$
     & $-4.6$ & $+0.003$ & $7.63$ & $5.18$ & $+0.3$ \\

$V_{7}$  & FP16 Training & \stages{\emptycell}{\modelcell}{\traincell}{\systemcell}{\emptycell}
     & $2.88$ & $4.49$ & \colorbox{SpringGreen}{$2.36$} & $3.73$ & $3.87$ & $3.46$ & $4.07$
     & $-20.5$ & $-0.014$ & $6.61$ & $4.21$ & $-18.5$ \\

$V_{8}$  & Sequence-Length Trim & \stages{\datacell}{\emptycell}{\traincell}{\systemcell}{\emptycell}
     & $8.61$ & \colorbox{Salmon}{$3,530$} & $2.95$ & $2.48$ & $4.01$ & $2.16$ & $2.73$
     & $-46.6$ & $-0.38$ & $4.44$ & $2.55$ & $-50.6$ \\

$V_{9}$  & Inference Engine & \stages{\emptycell}{\emptycell}{\emptycell}{\systemcell}{\infercell}
     & $3.31$ & $5.5$ & $2.92$ & $4.66$ & \colorbox{Salmon}{$112.9$} & \colorbox{SpringGreen}{$0.39$} & $4.71$
     & $-8.11$ & $-0.18$ & $7.44$ & $0.55$ & $-89.2$ \\

$V_{10}$ & Data-Loader Pin-Mem & \stages{\datacell}{\emptycell}{\traincell}{\emptycell}{\emptycell}
     & $9.61$ & $4.93$ & $2.43$ & $3.73$ & $3.60$ & $3.49$ & $4.08$
     & $-20.4$ & $-0.02$ & $6.64$ & $4.26$ & $-17.6$ \\

$V_{11}$ & Torch Compile & \stages{\emptycell}{\emptycell}{\traincell}{\systemcell}{\infercell}
     & $5.58$ & $4.43$ & $2.40$ & $2.77$ & $4.10$ & $3.81$ & $3.15$
     & $-38.4$ & $-0.01$ & $4.76$ & $4.45$ & $-13.9$ \\

$V_{12}$ & Attention Optimization & \stages{\emptycell}{\modelcell}{\traincell}{\emptycell}{\infercell}
     & $5.92$ & $4.91$ & $2.38$ & $3.46$ & $4.03$ & $2.87$ & $3.75$
     & $-26.8$ & $+0.04$ & $5.8$ & $3.39$ & $-34.5$ \\

$V_{13}$ & Layer Pruning (4 Top) & \stages{\emptycell}{\modelcell}{\traincell}{\systemcell}{\infercell}
     & $2.91$ & $5.31$ & $2.57$ & $3.98$ & $3.30$ & $3.70$ & $4.35$
     & $-15.0$ & $+0.10$ & $6.61$ & $4.30$ & $-16.7$ \\

$V_{14}$ & Layer Pruning (4 Bottom) & \stages{\emptycell}{\modelcell}{\traincell}{\systemcell}{\infercell}
     & $8.55$ & $4.72$ & $2.64$ & $3.98$ & $3.57$ & $3.68$ & $4.35$
     & $-15.2$ & $-0.10$ & $6.60$ & $4.27$ & $-17.3$ \\

$V_{15}$ & Layer Pruning (8 Top) & \stages{\emptycell}{\modelcell}{\traincell}{\systemcell}{\infercell}
     & $7.79$ & $4.80$ & $2.48$ & $3.10$ & $2.70$ & $2.94$ & $3.39$
     & $-33.8$ & $+0.072$ & $5.3$ & $3.41$ & $-34.1$ \\

$V_{16}$ & Layer Pruning (8 Bottom) & \stages{\emptycell}{\modelcell}{\traincell}{\systemcell}{\infercell}
     & $5.37$ & $5.23$ & $2.50$ & $3.10$ & $2.71$ & $2.92$ & $3.39$
     & $-33.8$ & $-0.11$ & $5.3$ & $3.37$ & $-34.8$ \\

$V_{17}$ & Layer Pruning (12 Top) & \stages{\emptycell}{\modelcell}{\traincell}{\systemcell}{\infercell}
     & $2.98$ & $4.82$ & $2.55$ & $2.40$ & $2.74$ & $2.18$ & $2.62$
     & $-48.9$ & $+0.12$ & $3.94$ & $2.51$ & $-51.4$ \\

$V_{18}$ & Layer Pruning (12 Bottom) & \stages{\emptycell}{\modelcell}{\traincell}{\systemcell}{\infercell}
     & $11.1$ & $4.89$ & $2.56$ & $2.40$ & $2.67$ & $2.19$ & $2.62$
     & $-48.8$ & $-0.04$ & $3.95$ & $2.52$ & $-51.2$ \\

$V_{19}$ & Layer Pruning (16 Top) & \stages{\emptycell}{\modelcell}{\traincell}{\systemcell}{\infercell}
     & \colorbox{Salmon}{$11.4$} & $4.44$ & $2.59$ & $1.48$ & $2.43$ & $1.43$ & $1.62$
     & $-68.3$ & $+0.11$ & $2.55$ & $1.63$ & $-68.5$ \\

$V_{20}$ & Layer Pruning (16 Bottom) & \stages{\emptycell}{\modelcell}{\traincell}{\systemcell}{\infercell}
     & $3.21$ & $5.14$ & $2.51$ & $1.47$ & $2.38$ & $1.42$ & $1.62$
     & $-68.4$ & $+0.05$ & $2.54$ & $1.61$ & $-68.9$ \\

$V_{21}$ & Layer Pruning (20 Top) & \stages{\emptycell}{\modelcell}{\traincell}{\systemcell}{\infercell}
     & $3.11$ & $4.98$ & $2.50$ & $0.73$ & $1.84$ & $0.65$ & \colorbox{SpringGreen}{$0.79$}
     & $-84.6$ & $-0.08$ & $1.14$ & $0.74$ & $-85.6$ \\

$V_{22}$ & Layer Pruning (20 Bottom) & \stages{\emptycell}{\modelcell}{\traincell}{\systemcell}{\infercell}
     & $3.65$ & $4.90$ & $2.51$ & $0.73$ & $1.60$ & $0.65$ & $0.80$
     & $-84.5$ & $+0.09$ & $1.14$ & $0.747$ & $-85.5$ \\
\midrule
$V_{23}$ & Attn+Pin-Mem+Opt8+GradAcc & \stages{\datacell}{\modelcell}{\traincell}{\systemcell}{\infercell}
     & $3.39$ & $4.54$ & $2.37$ & $4.09$ & $4.21$ & $3.72$ & $4.46$
     & $-12.9$ & $+0.11$ & $6.5$ & $4.08$ & $-21.0$ \\

$V_{24}$ & Inf+GradCkpt+LoRA+FP16 & \stages{\emptycell}{\modelcell}{\traincell}{\systemcell}{\infercell}
     & $6.24$ & $4.53$ & $3.19$ & $0.64$ & $107.0$ & $0.37$ & $0.69$
     & $-86.6$ & $-4.05$ & $1.69$ & $0.54$ & $-89.5$ \\

$V_{25}$ & GradAcc+FP16+Checkpoint & \stages{\emptycell}{\modelcell}{\traincell}{\systemcell}{\emptycell}
     & $4.89$ & $4.38$ & $2.36$ & $6.44$ & $3.97$ & $4.50$ & $6.89$
     & $+34.4$ & $+0.12$ & $10.3$ & $4.96$ & $-4.0$ \\

$V_{26}$ & Prune12B+Seq-Len+Compile & \stages{\datacell}{\modelcell}{\traincell}{\systemcell}{\infercell}
     & $5.75$ & $551$ & $2.67$ & $0.89$ & $2.59$ & $0.92$ & $0.99$
     & $-80.8$ & $-0.26$ & $1.45$ & $1.21$ & $-76.6$ \\

$V_{27}$ & Torch Compile + FP16 & \stages{\emptycell}{\modelcell}{\traincell}{\systemcell}{\infercell}
     & $3.31$ & $4.49$ & $2.44$ & $2.41$ & $4.28$ & $3.56$ & $2.77$
     & $-46.0$ & $+0.03$ & $4.1$ & $4.18$ & $-19.1$ \\

$V_{28}$ & Prune12B + Compile + FP16 & \stages{\emptycell}{\modelcell}{\traincell}{\systemcell}{\infercell}
     & $5.02$ & $4.59$ & $2.56$ & $1.17$ & $2.67$ & $1.74$ & $1.35$
     & $-73.7$ & $-0.07$ & $2.1$ & $2.10$ & $-59.3$ \\
$V_{29}$ & Attn+Pin-Mem+Opt8 & \stages{\datacell}{\modelcell}{\traincell}{\systemcell}{\infercell}
     & $2.73$ & $6.20$ & $2.70$ & $3.26$ & $3.64$ & $2.90$ & $3.56$
     & $-30.52$ & $-3.29$ & $5.53$ & $3.42$ & $-33.72$ \\
\midrule
$V_{30}$ & Optimal & \stages{\datacell}{\modelcell}{\traincell}{\systemcell}{\infercell}
     & $9.3$ & $4.3$ & $372.1$ & $0.20$ & $247.8$ & $0.25$ & $0.27$
     & \colorbox{LimeGreen}{$\textbf{-94.6}$} & $-4.0$ & \colorbox{LimeGreen}{$\textbf{0.54}$} & $0.37$ & \colorbox{LimeGreen}{$\textbf{-92.69}$} \\

\bottomrule
\end{tabularx}
\end{table*}

We define several experimental variants and categorize them into two families to answer two key practitioner questions: ``What is the quickest initial optimization step?'' and ``How far can we push energy efficiency once performance-preserving
gains have been exhausted?''
The first family of experiments consists of single-knob variants, each enabling exactly one optimization technique\textemdash 22 variants in total\textemdash allowing clear isolation and quantification of each technique’s direct impact. The second family comprises multi-knob bundles\textemdash six compound configurations that integrate orthogonal optimizations (\eg{} layer pruning combined with FP16 precision and Torch Compile). This approach tests our hypothesis that complementary techniques create synergistic efficiency improvements when employed together, thereby establishing a clear pathway toward maximum sustainable efficiency.

\noindent
\paragraph*{Experimental Setup}
All experiments involve fine-tuning ModernBERT-base on the BigVul dataset, a vulnerability-classification task comprising $217,000$ training samples. For accurate and comparable energy measurement, each variant is run three times on a clean and stable system, with no extraneous processes running before each run. Experiments are executed on a NVIDIA RTX 5000 Ada GPU (32 GB) paired with an AMD EPYC 9554P CPU. Energy consumption is monitored at one-second intervals for each process using 
CodeCarbon~\cite{benoit_courty_2024_11171501}. We fix all training hyperparameters, with only the relevant optimization configuration changed in each variant. The replication package is available on GitHub~\cite{Rajput_How_to_Tune_2025}.

\subsection{Knob Isolation Effects}
\label{sec:knob-isolation}

To understand the isolated impact of each optimization \emph{knob}, we activated them individually on the ModernBERT baseline (\vindex{0}, $0.512$ kWh, $0.994$ F1), resulting in 22 distinct variants. Table~\ref{tab:comprehensive-analysis} summarizes these results, highlighting several critical insights listed below.

\begin{itemize}
\item
\textbf{Training is the dominant energy consumer.}
Across all variants (\vindex{1}--\vindex{22}), the training stage consistently accounts for approximately 90--92\% of total energy consumption. Even aggressive pruning, such as removing 20 layers in \vindex{21}--\vindex{22}, leaves training as the primary energy sink, confirming it as the highest priority target for initial optimization.
\item
\textbf{Structural sparsity yields significant efficiency gains.}
Layer pruning stands out among performance-preserving
optimization techniques, providing substantial energy savings with minimal impact on model performance. Variants \vindex{21} (20-layer top pruning) and \vindex{22} (20-layer bottom pruning) achieve energy reductions of 84.6\% and 84.5\%, respectively, with negligible accuracy trade-offs ($\Delta$ F1: $-0.08$\% and $+0.09$\%). The energy savings from the pruning scale nearly linearly with the number of layers removed, reaching an optimal trade-off at 16-layer pruning (\vindex{19}, \vindex{20}: energy reduction of 68.4\%, $\Delta$ F1 $<$ $0.11$\%).

\item
\textbf{Compiler and precision optimizations provide performance-preserving efficiency wins.}
Techniques such as \texttt{torch.compile} (\vindex{11}) significantly reduces energy by 38.4\% and wall-time by approximately 40\%, with only a negligible accuracy impact ($\Delta$ F1: $-0.01$\%). Similarly, FP16 training (\vindex{7}) delivers a 20.5\% energy reduction while shortening runtime $16.6$\%, further emphasizing the practical benefits of these low-overhead methods. Sequence-length trimming (\vindex{8}) halves padding overhead, trimming energy $46.6$\% and time $43.9$\% at a moderate –0.38\% F1 cost.  LoRA (\vindex{2}) is frugal with (–$85.3$\% energy, –$78.6$\% time) but sacrifices ~4 \% F1, making it suitable only when accuracy budgets allow.
\item
\textbf{Memory-focused optimizations can be counterproductive.}
Not all memory-centric optimizations improve overall energy efficiency. Gradient checkpointing (\vindex{1}) incurs an energy penalty of 30.7\%, primarily due to the computational overhead introduced by recomputation. Similarly, static power limiting (\vindex{5}), although intuitively beneficial, results in a 26\% energy increase coupled with a substantial latency rise of 168\%. Quantization (\vindex{3}) yields only modest energy savings (5\%) but significantly extends runtime ($46.9$\%), suggesting applicability around resource-constrained devices.
\end{itemize}

\begin{figure*}[h] 
\centering
\includegraphics[width=0.6\linewidth]{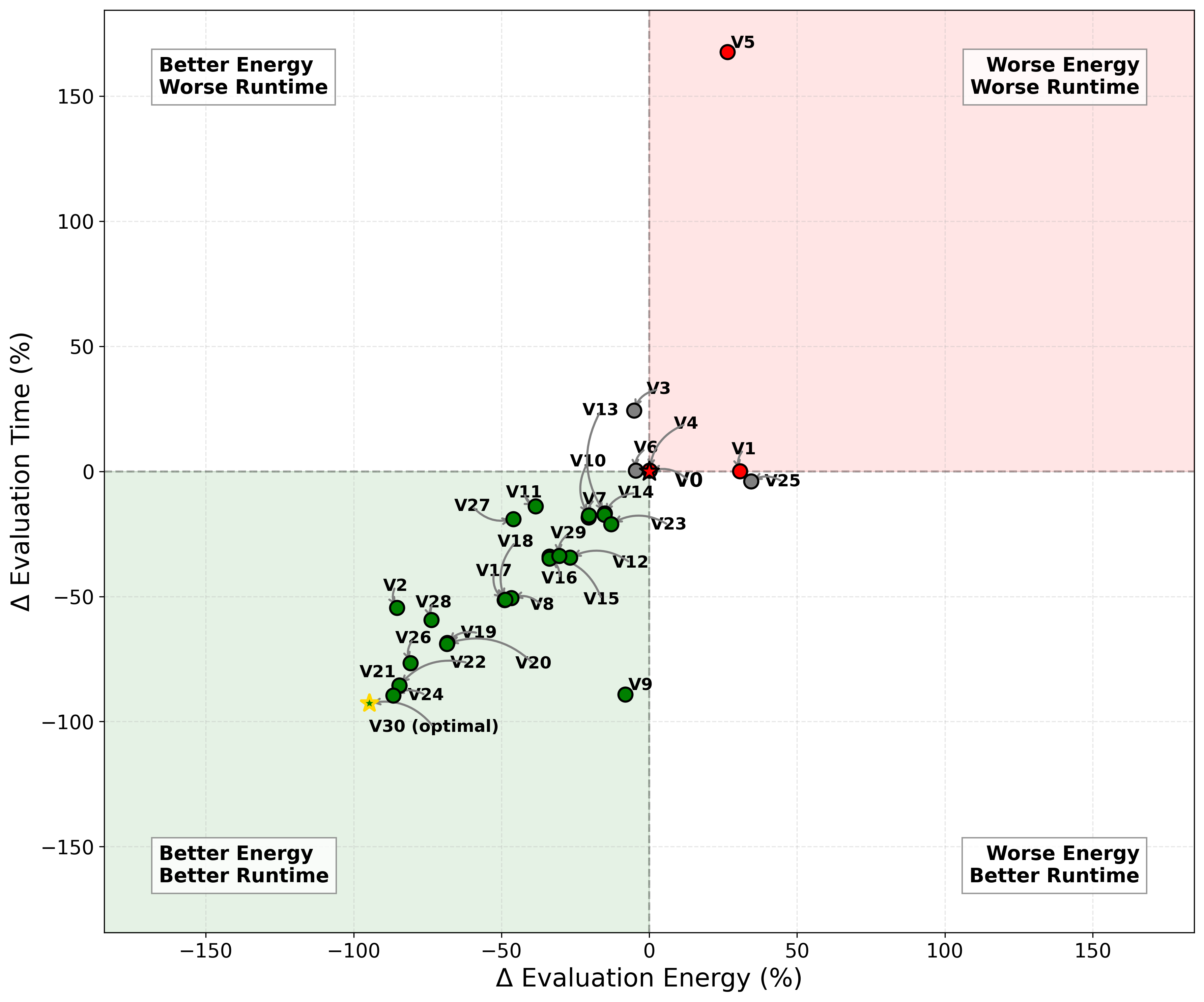}
\caption{Evaluation Energy vs. Evaluation Time Trade-off. The plot shows the percentage change in evaluation energy consumption ($\Delta$ Energy) versus the percentage change in evaluation time ($\Delta$ Time) relative to the baseline variant (\vindex{0}). Points in the bottom-left quadrant represent improvements in both energy efficiency and runtime, while points in the top-right quadrant indicate worse performance in both metrics. The baseline variant (\vindex{0}) is marked with a red star at the origin. Points are color-coded based on their performance relative to the baseline: green for improvements in both metrics, red for worse performance in both metrics and gray for mixed results.}
\label{fig:energy_time_quadrant}
\end{figure*}

Figure \ref{fig:energy_time_quadrant} provides practitioners with an actionable roadmap for optimization decisions by mapping the trade-off between evaluation time and energy consumption relative to the baseline configuration. This visualization serves three critical functions for production deployment:

\begin{enumerate}
    \item \textbf{Prioritization guidance}: Variants in the bottom-left quadrant (green) simultaneously reduce both energy and latency, representing accuracy-neutral first choices for immediate implementation (\eg{} layer pruning in \vindex{17}--\vindex{22}).
    
    \item \textbf{Tradeoff identification}: Upper-left quadrant variants reduce energy but increase runtime (\eg{} quantization in \vindex{3}), suitable for batch processing or edge deployments where latency is less critical than energy conservation.
    
    \item \textbf{Risk mitigation}: Right-half variants indicate counterproductive optimizations that degrade key metrics (\eg{} static power limiting in \vindex{1}, \vindex{4}, \vindex{5}), serving as warnings to avoid these configurations.
\end{enumerate}

Practitioners can leverage this visualization by first implementing bottom-left quadrant optimizations as foundational efficiency improvements, then progressively stacking additional knobs along favorable trajectories to amplify gains. When evaluating new configurations, teams should benchmark against \vindex{30}'s empirical lower bound (marked by the yellow star) to assess optimization potential. Finally, optimization efforts should conclude once application-specific accuracy or latency budgets are reached, ensuring resource-efficient deployment without unnecessary tuning.

\begin{rqbox}
\textbf{Takeaways:} Structural pruning consistently provides the highest single-knob energy savings. Memory-saving techniques that introduce additional computation often backfire, increasing total energy consumption. Static power limiting should be replaced by workload-aware methods or precision optimization to avoid increased latency and energy use.
\end{rqbox}

\subsection{Knob Combination Dynamics}
\label{sec:knob-combination}

Next, we evaluate seven multi-knob bundles (\textbf{\vindex{23}--\vindex{29}}) along with an optimal \textit{oracle} combination (\textbf{\vindex{30}}) to explore potential synergistic and antagonistic effects from combined optimizations. The results, that we summarize below, support our hypothesis that strategically curated optimizations can lead to significant cascading efficiency gains. 

\begin{itemize}
    \item 
\textbf{Synergistic combinations amplify efficiency.}
Combinations of orthogonal knobs\textemdash
those targeting distinct resource constraints such as computational intensity, arithmetic precision, and compiler optimization—consistently yield energy savings beyond the sum of their isolated effects. Specifically,
\begin{itemize}
\item \textit{pruning + sequence-length trim + torch compile (\vindex{26})} achieves an $80.8$\% energy reduction with a minor accuracy penalty ($\Delta$F1: -0.26\%). This demonstrates a compounded efficiency gain surpassing individual knob savings.
\item \textit{pruning + torch compile + FP16 (\vindex{28})} realizes a significant $73.7$\% energy saving while maintaining model accuracy with minimal degradation ($\Delta$F1: -0.07\%).
\item \textit{torch compile + FP16 (\vindex{27})} delivers a notable $46.0$\% energy reduction and slightly improves accuracy ($\Delta$F1: +0.03\%).
\end{itemize}

\item 
\textbf{Antagonistic combinations reduce or reverse gains.}
Conversely, combining optimizations that target similar resource bottlenecks often results in diminished or even negative efficiency outcomes. Specifically,
\begin{itemize}
\item \textit{gradient checkpointing + FP16 + gradient accumulation (\vindex{25})} unexpectedly increases energy consumption by $34.4$\%, primarily due to excessive recomputation overhead.
\item \textit{attention optimization + pin-memory + optimizer tuning (\vindex{29})} achieves a modest energy saving ($30.5$\%) but incurs substantial accuracy loss ($\Delta$F1: -3.29\%), highlighting a poor performance-energy trade-off.
\end{itemize}

\item 
\textbf{Stage sequencing amplifies cascading efficiencies.}
Early-stage optimizations such as sequence-length trimming (\vindex{8}) significantly reduce the downstream computational load, magnifying the overall efficiency gains in later stages. For example, combining sequence-length trimming with pruning and compilation (\vindex{26}) redistributes energy consumption across pipeline stages, resulting in significant reduction in total system energy.

To demonstrate the combination of the possible optimization techniques in a single variant, we combine the optimal strategies across stages into a single variant ($V_{30}$). The \textit{oracle} \vindex{30} achieves $94.6$\% reduction in energy, and
 $93.2$\% reduction in time, demonstrating the maximum potential efficiency gains observable in our study. Despite a $4$\% F1 score trade-off, this result conclusively demonstrates that strategically curated optimization knobs applied across stages can yield near-total energy savings while preserving core functionality.

\item 
\textbf{Optimization Strategies Along the Efficiency Frontier 
}
The energy-time tradeoff visualization in Figure~\ref{fig:energy_time_quadrant} reveals three pathways for practitioners to navigate efficiency optimizations:
\begin{itemize}
\item 
\textit{Accuracy-sensitive pathway:} Layer pruning (16–20 layers) combined with attention optimization delivers substantial energy savings ($68$\%–$85$\%) with negligible accuracy loss ($\Delta$F1 $<$ $0.11$\%). This approach occupies the desirable bottom-left quadrant in Figure~\ref{fig:energy_time_quadrant}, simultaneously reducing both energy and latency. 
\item 
\textit{Maximum-efficiency pathway:} Combining aggressive pruning, FP16 precision, and compiler optimizations achieves maximum energy savings ($73$\%–$94$\%) with modest to moderate accuracy trade-offs ($\Delta$F1: $-0.07$\% to $-4.00$\%).

\item
\textit{A Middle Ground:} Inference engine optimization (V9) complements both these pathway well, delivering dramatic time reductions ($89.2$\%) with minimal accuracy impact ($0.18$\% F1 reduction), making the combination ideal for latency-sensitive applications where performance preservation is critical. It give's modest energy gains ($-8.11$\%); however, if combined with variants from other stages as demonstrated in \vindex{24} and \vindex{30} it achieves up to 92.69\% time reduction, along with upto 94.6\% energy savings. This pathway pushes toward the empirical lower bound (yellow star in Figure~\ref{fig:energy_time_quadrant}), making it suitable for large-scale deployments where aggregate resource savings outweigh marginal performance differences.
\end{itemize}
\end{itemize}

\begin{rqbox}
\textbf{Key takeaways:} Combining orthogonal optimization knobs significantly amplifies energy efficiency. Avoid combining similar resource-targeted knobs, as it often negates potential gains. Sequence early-stage optimizations strategically to create cascading efficiency improvements downstream.
\end{rqbox}

\subsection{Stakeholder Responsibilities}
The $94.6$\% energy reduction (preserving $95.95$\% F1) underscores energy-efficiency's critical role in \aim{} pipelines, with implications for stakeholders:

\textbf{Industry:} Must shift from fragmented performance-centric development to holistic sustainable practices. Leader-boards currently incentivize marginal performance gains while ignoring environmental costs. Industry should adopt multi-faceted optimization combining techniques across data, model, training, system, and inference stages for cascading efficiency gains. A \textsc{cop}$29$-modeled consortium~\cite{cop29} should establish standardized sustainability metrics and carbon credits, alongside transparent energy reporting via public dashboards, third-party audits, and open datasets.

\textbf{Research:} Should expand Green \aim{} tools with new orthogonal optimization techniques, combinations, and energy-aware architectures. Must create standardized evaluation frameworks treating energy efficiency as a first-class metric alongside accuracy, including energy-aware principles, automated monitoring tools, and leader-boards rewarding efficient designs.

\textbf{Government:} Must incentivize sustainability through economic/regulatory measures, including \aim{}-specific carbon pricing and energy-based models reflecting environmental costs. Regulations should mandate efficiency standards, energy caps for large models, standardized impact reporting, and progressive renewable targets while promoting innovation and accountability.

\section{Final Remarks}
\aim{}'s rapid advancement brings both extraordinary opportunities and environmental challenges. Our experiments show that strategically combining orthogonal optimization techniques across the AI pipeline achieves $94.6$\% energy reduction while preserving $95.95$\% F1 score. This holistic approach makes energy efficiency a core design principle, creating cascading savings. Success requires coordinated action: industry implementing multi-knob optimization, researchers developing energy-aware architectures and evaluation frameworks, and policymakers establishing supportive regulations. Through conscientious collaboration, we can align \aim{} advancement with environmental sustainability. Treating sustainability as a catalyst (not a barrier) lets us harness \aim{}'s potential while minimizing its footprint.

\bibliographystyle{ieeetr}
\bibliography{references}
\balance{}

\end{document}